\documentclass[preprint,aps,prl,showpacs]{revtex4}

\begin{document}

\draft

\title{Quantum spiral bandwidth of entangled two-photon states}

\author{J. P. Torres, A. Alexandrescu and Lluis Torner}
\affiliation{ICFO-Institut de Ciencies Fotoniques, and Department
of Signal Theory and Communications, Universitat Politecnica de
Catalunya, 08034 Barcelona, Spain}

\begin{abstract}
{We put forward the concept of quantum spiral bandwidth of the
spatial mode function of the two-photon entangled state generated
in spontaneous parametric down-conversion. We obtain the bandwidth
using the eigenstates of the orbital angular momentum of the
biphoton states, and reveal its dependence with the length of the
down-converting crystals and waist of the pump beam. The
connection between the quantum spiral bandwidth and the entropy of
entanglement of the quantum state is discussed.}
\end{abstract}


\maketitle

\newpage

Entanglement, one of the most genuine features of quantum
mechanics, is a basic ingredient in quantum cryptography,
computing and teleportation \cite{nielsen1,zeilinger1}.
Spontaneous parametric down-conversion, namely the generation of
two lower frequency photons when a strong pump field interacts
with a nonlinear crystal, is a reliable source of entangled
photons. The generated two-photon state is entangled in
polarization \cite{kwiat1}, and most applications of parametric
down-conversion in quantum systems make use of such
spin-entanglement \cite{bouwmeester1}-\cite{jennewein1}. However,
entanglement is also embedded in the spatial mode functions that
describe the two-photon states \cite{photons}. Such spatial
entanglement occurs in an infinite dimensional Hilbert space and
is gaining increasing attention as a powerful way to encode and to
exploit quantum
information.\cite{pittman1,banaszek1,simonPRL02,navez1,gatti1,dawesPRA03}
For example, knowledge of the spatial structure acquired by the
mode functions of entangled signal and idler photons forms the
basis of multidimensional quantum imaging, and it can be used to
increase the efficiency of multidimensional quantum communication
protocols. The spatial structure of the photon states can be
expressed by a mode decomposition of their mode function in an
appropriate basis. The amount of entanglement of a quantum state
is directly related to the width of such modal expansion,
hereafter referred to as {\em quantum spatial bandwidth\/}. A
two-photon state described by a single mode is a product state,
while a quantum state described by an equi-distributed multimode
expansion corresponds to a maximally entangled state. Therefore, a
fundamental question that arises is how to act on the spatial
quantum distribution of a given mode expansion, hence how the
corresponding quantum information can be, e.g., {\it
concentrated}.

It has been recently observed experimentally that the photon pairs
generated in spontaneous down-conversion are entangled in orbital
angular momentum (OAM) \cite{mair1}. Such OAM entanglement allows
the implementation of arbitrary $d$-dimensional quantum channels
\cite{molina1}, thus it has been used to demonstrate violation of
Bell inequalities with qutrits \cite{vaziri1}. The corresponding
mode functions are naturally expressed in terms of eigenstates of
the paraxial OAM operator, whose spiral or winding topological
structure can be resolved experimentally using combinations of
holographic and filtering techniques. Therefore, such base
provides a powerful tool to explore the concept of spatial
multimode entanglement. In this Letter we put forward the full
characterization of the entangled photon pairs in terms of
eigenstates of the OAM operator and reveal how the corresponding
{\it quantum spiral bandwidth\/} depends on the shape of the beam
that pumps the down-converting crystal, and on the material
properties and length on the crystal. We obtain in analytical form
important similarity rules that hold for arbitrary input and
crystal conditions.

Let a quadratic nonlinear crystal of length $L$ be illuminated by
a quasi-monochromatic laser pump beam propagating in the $z$
direction. The spatial distribution of the pump beam writes
$E_p(x,z,t)=E_0(x)\exp \left[ i \left(\omega_p t-k_p z \right) \right]+c.c$,
where $\omega_p$ is the angular frequency of the pump
beam, $k_p$ is the wave number, $x$ is the position in the
transverse plane and $E_0$ is the transverse spatial field profile
at the input face of the crystal. The length of the crystal is
assumed to be much smaller than the diffraction length ($L_d$) of
the pump beam ($L << L_d$), so that the spatial distribution of
the pump beam in the transverse plane is independent of $z$. The
signal and idler photons are assumed to be monochromatic, with
$\omega_p=\omega_s+\omega_i$, where and $\omega_{s}$ and
$\omega_{i}$ are the frequency of the signal and idler photons,
respectively. This is justified by the use of narrow band
interference filters in front of the detectors.

In the paraxial approximation, the spin and the OAM can be
considered separately \cite{barnett1}. Under these conditions,
photons described by a mode function that is a Laguerre-Gaussian
(LG$^l_p$) mode are eigenstates of the OAM operator with
eigenvalue $l \hbar$ \cite{allen1}. The index $p$ is the number of
non-axial radial nodes of the mode and the index $l$, referred to
as the topological winding number, describes the helical structure
of the wave front around a wave front singularity or dislocation.
State vectors which are represented by a superposition of LG modes
correspond to photons in a superposition state, with the weights
of the quantum superposition dictated by the contribution of the
$l$-th angular harmonics. When the pump beam is a LG$^{l_0}_{p_0}$ mode,
under conditions of collinear phase-matching, the two-photon state
at the output of the nonlinear crystal can be written as a
coherent superposition of eigenstates of the OAM operator
\cite{franke1} that are correlated in OAM, i.e $l_1+l_2=l_0$,
where $l_1$ and $l_2$ refer to the OAM eigenvalues for the signal
and idler photons.  A photon state described by a LG mode can be
written as
\begin{equation}
\label{lg} |l p>= \int dq LG_{p}^{l}(q) a^{\dagger}(q)|0>,
\end{equation}
where the mode function in the spatial frequency domain writes
\begin{eqnarray}
& &  \label{laguerre} LG_{p}^{l}(\rho,\varphi)=\left( \frac{w_0^2 p!}{2 \pi (|l|+p)!} \right)^{1/2}
 \left( \frac{w_0 \rho_k}{\sqrt{2}} \right)^{|l|}
  L_{p}^{|l|} \left( \frac{\rho_k^2 w_0^2}{2} \right)
\exp \left( -\frac{\rho_k^2 w_0^2}{4} \right) \nonumber \\
& & \times \exp \left\{ i l \varphi_k+ i\left( p-\frac{|l|}{2} \right) \pi \right\},
\end{eqnarray}
with $\rho_k$ and $\varphi_k$ being the modulus and phase, respectively, of the
transverse coordinate $q$. The functions $L_p^l(\rho)$ are the
associated Laguerre polynomials and $w_0$ is the beam width.

The quantum state of the generated two-photon pair is given by
\cite{saleh1}
\begin{equation}
\label{state1} |\Psi>= \int dq_s dq_i \Phi(q_s,q_i)
a_s^{\dagger}(q_s) a_i^{\dagger}(q_i) |0,0>,
\end{equation}
where $q_{s,i}$ are the transverse coordinates in the spatial
frequency domain, $|0,0>$ is the vacuum state, and $a_s^{\dagger}$
and $a_i^{\dagger}$ are creation operators for the signal and
idler modes. Under conditions of collinear propagation of the
pump, signal and idler photons, the mode function $\Phi(q_s,q_i)$
can be written in the paraxial approximation \cite{walborn1} as
$\Phi(q_s,q_i)=E_0(q_s+q_i) W(q_s-q_i)$, where normalization of
the state requires $\int dq_s dq_i |\Phi(q_s,q_i)|^2=1$. Here
$E_0$ is the normalized field distribution of the pump beam in the spatial
frequency domain, and the function $W$, which comes from the phase
matching condition in the longitudinal direction, is given by
\begin{equation}
\label{state} W(q_s,q_i)= \sqrt{\frac{2L}{\pi^2 k_p}} sinc \left(
\frac{|q_s-q_i|^2 L}{4 k_p}\right) \exp \left(
-i\frac{|q_s-q_i|^2 L}{4 k_p} \right).
\end{equation}
One can decompose the quantum state  $|\Psi>$ in the base of the
eigenstates of the OAM operator, as
\begin{equation}
\label{state2}
 |\Psi>= \sum_{l_1,p_1} \sum_{l_2,p_2}
C_{p_1,p_2}^{l_1,l_2} |l_1,p_1;l_2,p_2>,
\end{equation}
where $(l_1,p_1)$ correspond to the signal mode, $(l_2,p_2)$
correspond to the idler mode and the amplitude $C_{p_1,p_2}^{l_1,l_2}$ writes
\begin{equation}
\label{amplitude1}
 C_{p_1,p_2}^{l_1,l_2} =\int dq_s dq_i
\Phi(q_s,q_i) \left[ LG_{p_1}^{l_1}(q_s) \right]^* \left[
LG_{p_2}^{l_2}(q_i) \right]^*.
\end{equation}
The weights of the quantum superposition are given by $P_{p_1,p_2}^{l_1,l_2}=|C_{p_1,p_2}^{l_1,l_2}|^2$,
which gives the value of the joint detection probability for finding one photon in the signal mode $(l_1,p_1)$
and one photon in the idler mode $(l_2,p_2)$.
The two-photon state can be also characterized by the amplitude
$A(x_1,z_1,t_1;x_2,z_2,t_2)=<0,0|E_1^{+}E_2^{+}|\Psi>$, at positions
$(x_1,z_1)$ and $(x_2,z_2)$ and times $t_1$ and $t_2$, where
$E^{+}(x,z,t)$ is the electric field operator. If one makes use of the paraxial 
approximation for the electric field operator of a photon travelling in vacuum \cite{saleh1,walborn1}, 
the amplitude can be written as
\begin{equation}
\label{amplitude2} A(x_1,z_1,t_1;x_2,z_2,t_2)= \sum_{l_1,p_1}
\sum_{l_2,p_2} C_{p_1,p_2}^{l_1,l_2} LG_{p_1}^{l_1}(x_1,z_1)
LG_{p_2}^{l_2}(x_2,z_2) \exp \left( -i\omega (t_1+t_2) \right).
\end{equation}
Such amplitude might be employed, e.g., to draw analogies between
the spatial structure of the two-photon quantum states and the
properties of the corresponding incoherent classical radiation
(see \cite{gatti1,saleh1}).

The amplitudes $C_{p_1,p_2}^{l_1,l_2}$ depends on two normalised
parameters: the normalised pump beam width
$\bar{w}_p=w_p/\sqrt{\lambda_p L}$ and the normalised beam width
of the LG modes, $\bar{w}_0=w_0/\sqrt{\lambda_p L}$, where
$\lambda_p$ is the wavelength of the pump beam in vacuum. The
assumption that the diffraction length ($L_d$) of the pump beam is
much smaller than the length of the crystal ($L<<L_d$), requires
the condition $\pi n_p \bar{w}_p >>1$ to be fulfilled, with $n_p$
being the refractive index of the pump beam inside the crystal.
Here we always consider $\bar{w}_p \geq 1$. For a typical value of
the pump wavelength $\lambda_p=0.4 \mu$m, and a crystal length of
$L=1$ mm, a pump beam width of $w_p\sim 20 \mu$m correspond to a
normalized value of $\bar{w}_p\sim 1$.

In Figure 1 we plot the contribution to the mode decomposition of
the quantum state $|\Psi>$ of all the harmonics with the same
value of the indexes $l_1$ and $l_2$, i.e.
$P_{l_1,l_2}=\sum_{p_1,p_2=0}^{\infty} P^{l_1,l_2}_{p_1,p_2}$. In
all cases cases the pump beam is a Gaussian mode. The distribution
$P_{l_1,l_2}$ depends only on the normalised width $\bar{w}_p$.
Thus, for a given material, the OAM distribution of the state
$|\Psi>$ depends on the ratio $w_p/\sqrt{\lambda_p L}$, a result
which gives an important scaling rule. In Figure 2(a) we show the
dependence of $P_{l_1,l_2}$ on $\bar{w}_p$ for different values of
$l_1$ and $l_2$, with $l_1+l_2=0$. One observes that the spiral
bandwidth increases with $\bar{w}_p$, therefore it can be made
larger by increasing the pump beam width or by decreasing the
crystal length. Notice that by doing so the coupling efficiency in
the detection state is also modified (see, e.g., \cite{bovino1}).
Although the total contribution $P_{l_1,l_2}$ of modes with a
given index $l_1$ and $l_2$ depends only on $\bar{w}_p$, the
actual amplitude distribution $C_{p_1,p_2}^{l_1,l_2}$ depends also
on $\bar{w}_0$. In Figure 2(b) we show the contributions of the
modes with different index $p_1$ and $p_2$ for $\bar{w}_p=1$ and
$l_1=l_2=0$, which is representative of a general case. We plot
$C_{p_{max}}^{0,0}=\sum_{p_1,p_2=0}^{p_{max}}
P_{p_1,p_2}^{l_1,l_2}$ as a function of $p_{max}$. The plot
reveals that there is an optimal value of $\bar{w}_0$ for which
the contribution of $P_{0,0}^{0,0}$ is maximum, and only a few
modes with different index $p$ make significant contributions to
the total weight. Away from the optimal $\bar{w}_0$ value, a large
number of modes are required to represent the quantum state.

In most applications that make use of the OAM of the photons, one
projects into a subspace of the complete Hilbert space that
describes the mode function of the photon (e.g., \cite{vaziri1}).
This implies considering only a fraction of the mode space. Let us
consider only modes with $p_1=p_2=0$, and thus thereafter we
define $|l_1,l_2> \equiv |l_1,p_1=0; l_2, p_2=0>$. In this
subspace, the two-photon state can be written as
\begin{equation}
\label{state3}
 |\Psi>= \sum_{l_1=-\infty}^{\infty} \sum_{l_2=-\infty}^{\infty}
C_{0,0}^{l_1,l_2} |l_1,l_2>.
\end{equation}
Since the amplitude $C_{0,0}^{l_1,l_2}$ depends on both normalised
parameters $\bar{w}_p$ and $\bar{w}_0$, the expansion given by Eq.
(\ref{state3}) depends on the base of LG modes considered. In
general, calculation of the amplitudes $C_{p_1,p_2}^{l_1,l_2}$
requires a $4$-dimensional integration in the spatial frequency
domain. However, for a pump beam with a spatial field profile
corresponding to a LG$_0^{l_0}$ mode, we were able to obtain the
value of the amplitude $C_{0,0}^{l_1,l_2}$ is analytic form.
Namely,
\begin{equation}
\label{formula1}
C^{l_1,l_2}_{0,0} =A_0 \Gamma \left( \frac{|l_0|+|l_1|+|l_2|}{2}+1
\right)   \tan^{-1} \frac{1}{i+\frac{w_0^2k_p}{2L}}
\end{equation}
for $l_1 \cdot l_2 < 0$, and
\begin{eqnarray}
\label{formula2}
& & C^{l_1,l_2}_{0,0}= A_0
\sum_{n=0}^{\min(|l_1|,|l_2|)}
 \frac{ |l_1|! |l_2|!}{(|l_1|-n)! (|l_2|-n)! \left( n! \right )^2}
\left( \frac{w_0^2}{2w_p^2+w_0^2}\right)^{-n}
\frac{\Gamma(\frac{|l_0|+|l_1|+|l_2|-2n}{2}+1)}{\left(1+i\frac{4L}{w_0^2 k_p}\right)^{\frac{n}{2}}} \nonumber \\
 & & \times \Gamma(n) \sin \left( n\tan^{-1}\frac{1}{i+\frac{w_0^2 k_p}{2L}} \right)
\end{eqnarray}
for $l_1 \cdot l_2 \ge 0$. In these expressions, $A_0$ stands for
\begin{equation}
\label{formula3}
 A_0= (-1)^{|l_0|-|l_1|-|l_2|} \left(\frac{k_p w_p^2}{\pi
L}\right)^\frac{1}{2} \left(\frac{w_p}{w_0}\right)^{|l_p|}
\left(\frac{2w_0^2}{2w_p^2+w_0^2}
\right)^{\frac{|l_0|+|l_1|+|l_2|}{2}+1}
\frac{2^{\frac{|l_0|-|l_1|-|l_2|}{2}+1}}{\sqrt{|l_0|! |l_1|!
|l_2|!}}.
\end{equation}
In both Eqs. (\ref{formula1}) and (\ref{formula2}), the relation
$l_0=l_1+l_2$ holds. It is worth stressing that Eqs.
(\ref{formula1})-(\ref{formula3}) provide the analytical
expression of the quantum state under general conditions in terms
of the pump beam and the crystal properties. Moreover, this expansion corresponds
to performing a Schmidt decomposition for continuos variables in this subspace, thus allowing to
calculate the entropy of entanglement.\cite{parker1}

 In Figure 3, we
plot the expansion in OAM eigenstates of the two-photon state for
pump beams with $l_0=0,1,2$ and $\bar{w_p}=\bar{w}_0=1$. In the
three cases represented, the subspace we are considering
represents more than $40\%$ of the corresponding complete Hilbert
space. The spiral bandwidth of the expansion in terms of states
$|l_1,l_2>$ increases with $l_0$. Equations (\ref{formula1}) and
(\ref{formula2}) show that, for a given pump beam, the phase of
the amplitude $C_{0,0}^{l_1,l_2}$ changes for each mode with $l_1
\cdot l_2 <0$, but it is the same for all modes with $l_1 \cdot
l_2 \geq 0$.

The concept can be extended to the general case of a pump beam
whose spatial field distribution is a coherent superposition of LG
modes, so that $E_0(x)=\sum_{m} C_{m} LG_0^m(x)$, with
$m=0,2,4..M$. The LG modes $LG_0^m(x)$ are the corresponding
expression in the spatial domain of the LG modes given by Eq.
(\ref{laguerre}). The energy flow $I$ of the pump beam can be
written as \cite{molina1} $I=2 \epsilon_0 c n_p \sum_{m}^{M}
|C_{m}|^2$. If we restrict to the subspace spanned by the states
of the form $|l,l>$, one generates an entangled two-photon state
of the form $|\Psi> \sim \sum_{n=0}^{N} \gamma_n |n,n>$. By
making use of Eqs. (\ref{formula1}) and (\ref{formula2}), one
obtains that the amplitudes $\gamma_n$ of the quantum state
generated when the crystal is pumped by the superposition of LG
modes are
\begin{equation}
\label{final}
\gamma_{n} = \frac{\sqrt{(2n)!}}{n!}
\left[   \frac{2\bar{w}_p}{\bar{w_0} \left( 2\bar{w}_p^2+1 \right)}
\right]^{2n+2} C_{2n},
\end{equation}
This expression reveals that properly tailoring the spatial
features of the beam that pumps the down-converting crystal allows
generation of optimal quantum states for different quantum
information protocols, such as the entanglement enhancement
required for obtaining maximally entangled states.\cite{bennett1}
One approach for the realization of entanglement concentration is
to perform mode filtering operations on the two-photon state that
is generated at the output of the crystal \cite{vaziri2}. In
contrast, Eq. (\ref{final}) shows that the required quantum state
could be obtained through appropriate engineering of the spatial
properties of the pump beam \cite{torres1}. It also shows that the
phase of the state $|n,n>$ corresponds to the phase of the
corresponding mode of the classical pump beam, $arg\{C_{2n}\}$, a
result consistent with the experimental observation by Ou et. al.,
that the down-converted photons carry information about the phase
of the pump beam \cite{ou1}.

In conclusion, we have obtained and analyzed the detailed quantum
spatial structure of the two-photon entangled states generated in
parametric down-conversion in terms of the eigenstates of the
orbital angular momentum operator. We put forward the related
concept of quantum spiral bandwidth, and showed its dependence on
the pump beam and down-converting crystal. This allows, e.g., to
define an effective finite Hilbert space where entanglement takes
place.\cite{law1} Engineering such quantum spatial bandwidth
should be a key tool to optimize multidimensional quantum
information schemes, for example to increase the resolution of
two-photon imaging using entangled photons (e.g.,
\cite{abouraddy1}) and to enhance the efficiency of relevant
multidimensional quantum cryptography protocols (e.g.,
\cite{peres1}).

\acknowledgments

This work was supported by the Generalitat de Catalunya and by the
Spanish Government through grant BFM2002-2861.

\newpage



\newpage
\normalsize

 \noindent {\bf Figure captions}

\noindent
{\bf Fig. 1}: Mode distributions $P_{l_1,l_2}$ for different values of the parameter $\bar{w}_p$. (a) $\bar{w}_p=1$, (b) $\bar{w}_p=2.5$ and (c) $\bar{w}_p=5$. The pump beam is a gaussian mode ($l_p=0$), so that $l_1+l_2=0$. The $x$-axis represents the value of $l_1$.

\noindent
{\bf Fig. 2}: (a) Value of the weight $P_{l_1,l_2}$  for $(l_1,l_2)=(0,0)$, $(1,-1)$ and $(2,-2)$, as a function of the normalized pump beam width $\bar{w}_p$. The labels show the value of $(l_1,l_2)$. The pump beam is a gaussian mode.
(b) Weight $C_{p_{max}}^{0,0}$ as a function of the index $p_{max}$ for different values of $\bar{w}_0$, as shown in the labels. In all cases $\bar{w}_p=1$.

\noindent
{\bf Fig. 3}: Mode decomposition in the subspace $|l_1,p_1=0; l_2,p_2=0>$ for several pump beam. In (a), (c) and (e) we plot the weight $P_{0,0}^{l_1,l_2}$ of the mode distribution , and in (b), (d) and (f) we plot the phase
$arg\{ C_{0,0}^{l_1,l_2} \}$ of each mode. (a), (b): $l_0=0$; (c), (d): $l_0=1$ and (e), (f): $l_0=2$.

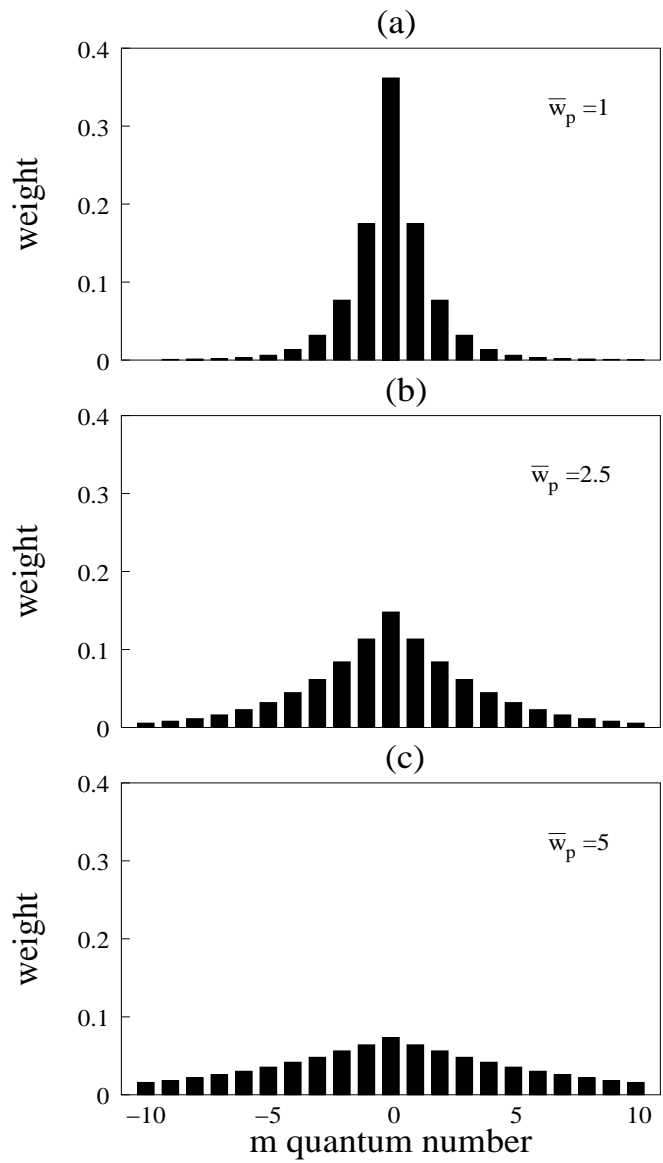
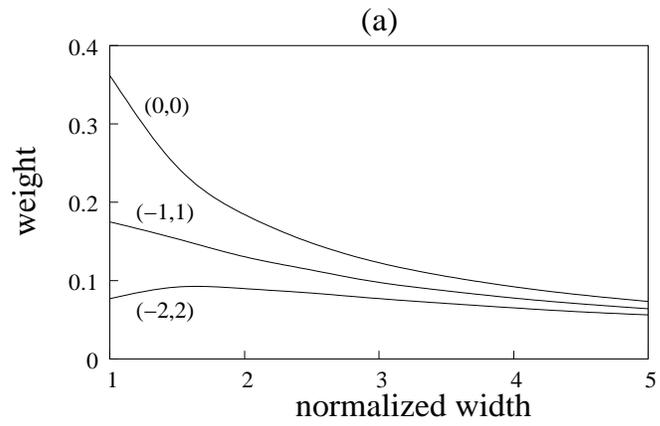
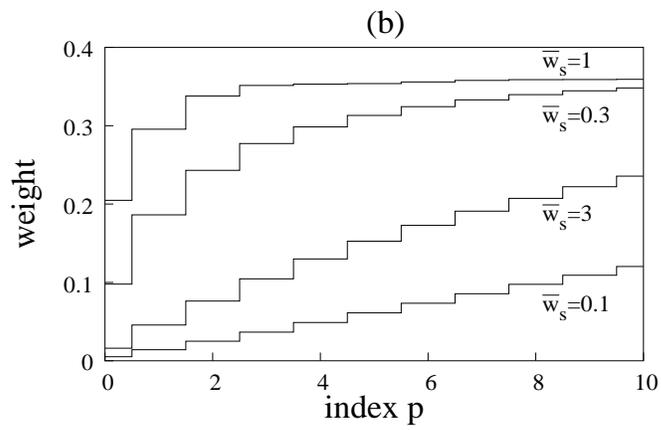
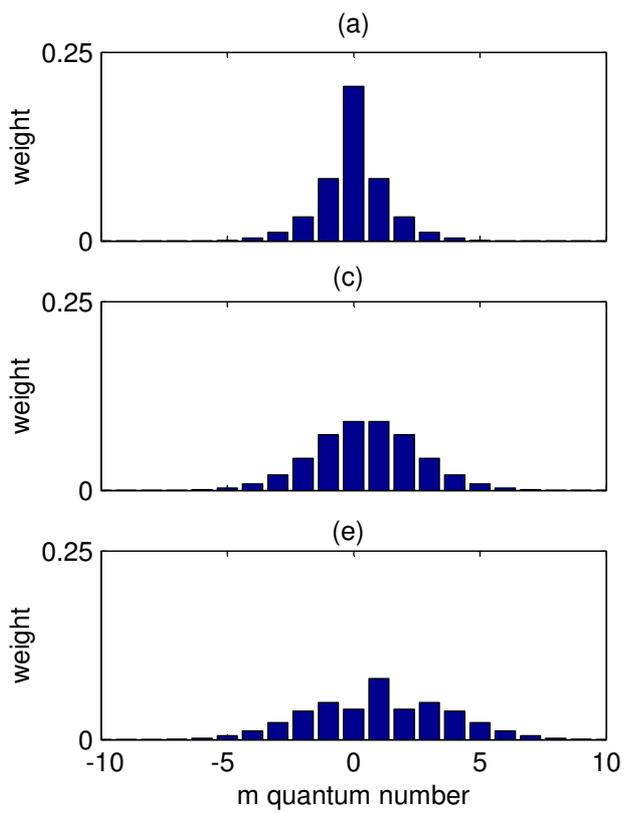
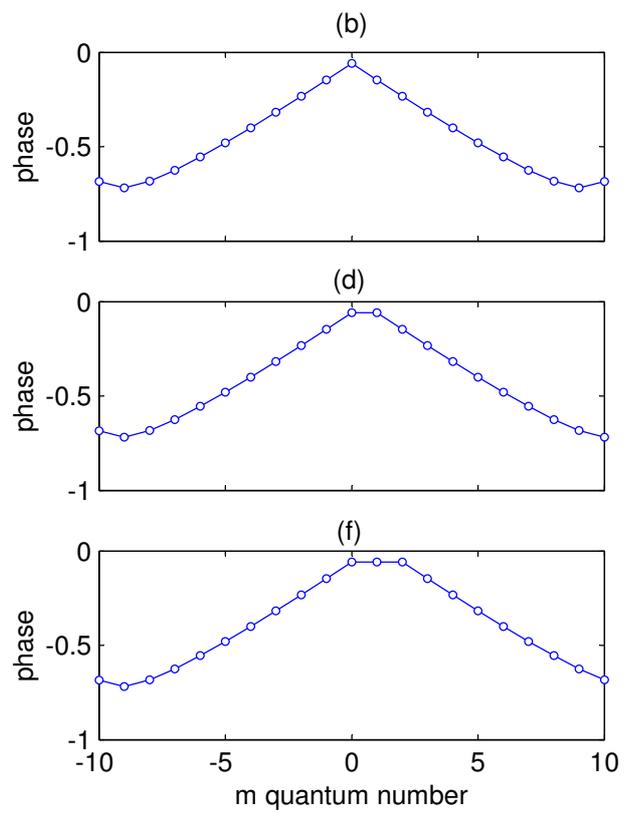
\end{document}